\documentclass{pasj00}

\begin{document}
\SetRunningHead{T.\ Kodama et al.}{Panoramic Views of Distant Clusters}
\Received{2004 December 4}
\Accepted{2005 February 21}

\def\gsim{\mathrel{\raise0.35ex\hbox{$\scriptstyle >$}\kern-0.6em 
\lower0.40ex\hbox{{$\scriptstyle \sim$}}}}
\def\lsim{\mathrel{\raise0.35ex\hbox{$\scriptstyle <$}\kern-0.6em 
\lower0.40ex\hbox{{$\scriptstyle \sim$}}}}
\def\halpha{{\rm H$\alpha$}}
\def\oii{{\rm [O{\sc ii}]}}
\def\ewoii{{\rm W$_\circ$([O{\sc ii}])}}
\def\kmsmpc{{\,\rm km\,s^{-1}Mpc^{-1}}} 
\def\kms {{\,\rm km\,s^{-1}}} 
\def\zphot{{\,$z_{\rm phot}$}}

\title
{Panoramic Views of Cluster-Scale Assemblies Explored by Subaru Wide-Field Imaging
\thanks
{Based in part on data collected at the Subaru Telescope, which is
operated by the National Astronomical Observatory of Japan.
Based also in part on observations obtained with XMM-Newton, an ESA science
mission with instruments and contributions
directly funded by ESA Member States and NASA, USA.}
}

\author{
Tadayuki \textsc{Kodama}\altaffilmark{1},\\
Masayuki \textsc{Tanaka}\altaffilmark{2},
Takayuki \textsc{Tamura}\altaffilmark{10},
Hideki \textsc{Yahagi}\altaffilmark{1},
Masahiro \textsc{Nagashima}\altaffilmark{7},
Ichi \textsc{Tanaka}\altaffilmark{5},\\
Nobuo \textsc{Arimoto}\altaffilmark{1},
Toshifumi \textsc{Futamase}\altaffilmark{5},
Masanori \textsc{Iye}\altaffilmark{1},
Yoshikazu \textsc{Karasawa}\altaffilmark{5},\\
Nobunari \textsc{Kashikawa}\altaffilmark{1},
Wataru \textsc{Kawasaki}\altaffilmark{9},
Tetsu \textsc{Kitayama}\altaffilmark{11},
Hideo \textsc{Matsuhara}\altaffilmark{10},\\
Fumiaki \textsc{Nakata}\altaffilmark{8},
Takaya \textsc{Ohashi}\altaffilmark{12},
Kouji \textsc{Ohta}\altaffilmark{6},
Takashi \textsc{Okamoto}\altaffilmark{1,8},\\
Sadanori \textsc{Okamura}\altaffilmark{2,4},
Kazuhiro \textsc{Shimasaku}\altaffilmark{2,4},
Yasushi \textsc{Suto}\altaffilmark{3,4},
Naoyuki \textsc{Tamura}\altaffilmark{8},\\
Keiichi \textsc{Umetsu}\altaffilmark{9},
and Toru \textsc{Yamada}\altaffilmark{1}
}
\altaffiltext{1}{National Astronomical Observatory of Japan, Mitaka, Tokyo 181--8588, Japan}
\email{kodama@optik.mtk.nao.ac.jp}
\altaffiltext{2}{Department of Astronomy, School of Science, University of Tokyo, Tokyo 113--0033}
\altaffiltext{3}{Department of Physics, School of Science, University of Tokyo, Tokyo 113--0033}
\altaffiltext{4}{Research Center for the Early Universe, School of Science, University of Tokyo, Tokyo 113--0033}
\altaffiltext{5}{Astronomical Institute, Tohoku University, Aoba-ku, Sendai 980--8578}
\altaffiltext{6}{Department of Astronomy, Kyoto University, Sakyo-ku, Kyoto 606--8502}
\altaffiltext{7}{Department of Physics, Kyoto University, Sakyo-ku, Kyoto 606--8502}
\altaffiltext{8}{Department of Physics, University of Durham, South Road, Durham DH1 3LE, UK}
\altaffiltext{9}{Institute of Astronomy and Astrophysics, Academia Sinica, Taipei 106, Taiwan}
\altaffiltext{10}{Institute of Space and Astronautical Science, Sagamihara, Kanagawa 229--8510}
\altaffiltext{11}{Department of Physics, Toho University, Funabashi, Chiba 274-8510}
\altaffiltext{12}{Department of Physics, Tokyo Metropolitan University, Hachioji, Tokyo 192-0397}


\maketitle

\begin{abstract}
We have started PISCES project; a panoramic imaging and spectroscopic survey of
distant clusters on Subaru.  It exploits the unique wide-field imaging
capability of Suprime-Cam, which provides a 34$'$$\times$27$'$ field of view
corresponding to a physical area of 16$\times$13~Mpc$^2$ at $z\sim1$.
We plan to target 15 clusters in total at $0.4\lsim z\lsim 1.3$.
In this paper, we report on our first results concerning the inner structures
and large-scale structures of two distant clusters at $z=$0.55 and 0.83
together with the earlier results on a $z=0.41$ cluster.
The photometric redshift technique has been applied to multi-color data
in order to efficiently remove most of the foreground/background galaxies
so as to isolate the cluster member candidates.
We have found large-scale filamentary structures around the clusters,
extending out to $>$5~Mpc from the cores, as well as complex inner structures.
The galaxy distributions in the inner regions of the clusters look
similar to the X-ray intensity maps, suggesting that most of the
optical structures trace physically bound systems.
We also compared the structures of the three clusters with those
of model clusters in a numerical simulation ($N$-body +
semi-analytic model) by parameterising the shapes of the
iso-density contours of galaxies, and found a broad agreement.
Our results provide good evidence that cluster-scale
assembly takes place along filaments during hierarchical clustering,
while the structures found here
need to be confirmed spectroscopically in the near future.
\end{abstract}

\section{Introduction}
\label{sec:intro}

Clusters of galaxies are the largest-scale objects in the Universe in which
dynamical relaxation has fairly advanced. They consist of 100--1000
galaxies, X-ray emitting hot plasma, and dark matter, which binds the systems.
Since the dynamical time scale of clusters is comparable to the Hubble time,
these systems are still presently dynamically evolving,
and thus are supposed to somehow keep memory of the
cosmological initial conditions.
For this reason, clusters of galaxies are often used as
cosmological probes at various wavelengths
(e.g., Kitayama et al.\ 1998).
Now that the cosmological parameters have been determined with a
reasonable accuracy by microwave background radiation and distant supernovae
(Spergel et al.\ 2003),
our primary motivation is now directed to understanding the major constituents
of the Universe, such as galaxies, baryonic gas, and dark matter, which are
playing major roles in clusters of galaxies.

According to the cold dark-matter scenario, small-scale objects collapsed
first as a result of gravitational instability, and then they pulled together
and merged one another to form bigger and bigger systems (dark-matter halos)
with time.
In each dark-matter halo the gas condensed and formed stars --- the birth of
shining galaxies.
During the course of assembly to the larger systems (groups/clusters),
galaxies experienced environmental effects by interactions with other
galaxies, cluster potential, and/or intra-cluster gas.  Hence, their
star-formation activities as well as morphologies are largely affected and
truncated (e.g., Larson et al.\ 1980;
Lavery, Henry 1986; Moore et al.\ 1996; Abadi et al.\ 1999;
Hashimoto, Oemler 2000).
In this way, galaxy evolution must be closely related to the structure
evolution of the Universe and the change of galaxy environment.
These environmental effects (a posteriori) are very important
because the well-known morphology--density relation
(Dressler 1980; Dressler et al.\ 1997) and the
origin of the Hubble sequence of galaxies are possibly related to
these processes.  These are additional effects on top of
the intrinsic effects (a priori) set by the cosmological initial conditions,
i.e., cluster cores started off from higher density peaks in the early
Universe, where galaxies formed quickly in a biased manner.
(e.g., Kaiser 1984; Bardeen et al.\ 1986; Cen, Ostriker 1993).

In spite of recent vigorous progress in this field, the central
physical processes behind the environmental effects still remain elusive.
The key possible processes that have been proposed include:
(1) ram-pressure stripping, a sudden removal of gas by the
interactions with the intra-cluster medium (ICM) (e.g., Abadi et al.\ 1999);
(2) tidal encounters, often accompanied by an intense star burst
(e.g., Lavery, Henry 1986; Hashimoto, Oemler 2000);
and (3) strangulation, a gradual consumption of the remaining disk gas
without any further gas supply from the halo (Larson et al.\ 1980).
Being limited by the light-collection power of telescopes and/or
narrow field coverage of detectors, observers have concentrated
on either cluster cores or general fields,
and have had little knowledge concering the transition regions that bridge
these two extreme environments.
We do not know the matter distribution (galaxy/dark-matter) there, either.

\subsection{PISCES Project}
\label{sec:pisces}

\subsubsection{Basic concepts}
\label{sec:concepts}

The advent of the Subaru Telescope (Iye et al.\ 2004)
with a unique combination of great light-collection power and 
large field of view,
has opened a new window in the study of distant clusters.
This has enabled us to look into a more distant Universe
($z\sim 1$) which has abundant information on galaxy/cluster evolution,
and to reach fainter objects ($>$$M^*$+1.5) where evolution is stronger.
At the same time, with its wide-field camera Suprime-Cam
(Miyazaki et al.\ 2002)
covering $30\arcmin$ (14~Mpc in physical scale or 29~Mpc in comoving scale
at $z=1$), we can view from cluster cores
through the transition region out to the general field, all at once.
Taking this unique advantage of Subaru,
we started the PISCES project (Panoramic Imaging and
Spectroscopy of Cluster Evolution with Subaru)
by combining several active research groups in Japan working
on clusters of galaxies in various aspects
(optical-NIR/X-ray/lensing/simulation).
This large collaborative project aims to carry out
systematic deep and wide observations of distant
clusters at various evolutionary stages, and compare
their fundamental physical quantities (mass, luminosity, colors,
spectral indices, morphology, and kinematics) in detail as a function of
the environment (structure, local density) and time (redshift).
This process will eventually prove when, where and how the galaxies form
and evolve.
The basic idea is to utilize clusters of galaxies as landmarks in the
distant Universe, and to map out the large-scale structures around them,
and reconstruct the history of galaxy evolution along with
the hierarchical structure formation of the Universe.
The critical issues we aim to address with PISCES are the following two:

\begin{itemize}
\item {\bf Cluster-scale assembly and the spatial bias:} 
We will map out the large-scale structure beyond the cluster cores
(e.g., West et al. 1998; Lubin et al. 2000) and trace the history of
cluster-scale assembly along the filaments.
By using the weak-lensing technique, we will also map the dark-matter
distribution. Galaxies do not necessarily trace the distribution of
dark matter faithfully (e.g., Kaiser 1984; Bardeen et al.\ 1986;
Cen, Ostriker 1993).
By quantifying this galaxy formation bias as a function of the redshift,
scale and density, we will place strong constraints on the galaxy formation
model.
\item {\bf Star formation history and the environmental effect:}
By comparing the galaxy properties as a function of environment and time
with the aid of a population synthesis technique, we will derive
the `environment-dependent' star-formation rate, mass, and morphology of
galaxies.
By watching when, where, and how the environmental effects exert on galaxies,
we will identify the major physical processes involved in morphological
transformation, and reveal the origin of the Hubble sequence of galaxies.
\end{itemize}

\subsubsection{Cluster sample}
\label{sec:sample}

The total planned targets of our PISCES project are 15 X-ray detected
clusters at 0.4$<$$z$$<$1.3,
covering various evolutionary stages ($z$) and richness ($L_{\rm X}$),
in coordination with XMM/Chandra and HST/ACS observations
(see table~\ref{tab:targets}).

The criteria of our target selection are the following:
\begin{itemize}
\item X-ray detection: We select clusters that are detected in X-rays.
X-ray selection is less prone to the projection effect along the line of sight
than optical selection.
Although we may be biased towards the densest regions at all epochs, this is
not very problematic in a time-sequence comparison, since the wide-field
imaging will cover all environments from the densest cluster cores to the
nearby groups and to the low-density filaments connecting the clumps.
\item Redshift range: We limit the cluster sample to the redshift range of
$0.4\lsim z\lsim 1.3$, so that the Suprime-Cam filters alone in the optical
($BVRi'z'$) can neatly cover the 4000~\AA\, break feature in the rest-frame,
and hence the photometric redshift technique works well (subsection 2.2).
\item Redundancy:
We select multiple (3--5) clusters in each of the
four redshift ranges ($z\sim0.4$, 0.55, 0.85, and 1.2).
Clusters of galaxies are statistical objects by nature, and are likely to
have large varieties; it is thus important to average the properties
over many clusters of different richness at a given epoch to obtain
general views.
\item Coordination with other projects:
We include many common targets with other major cluster projects,
especially with XMM/Chandra and HST/ACS observations
(e.g., Blakeslee 2001, private communication).
Such coordination is important, since diffuse X-ray emission traces the
distribution of gravitationally bound systems and provides the properties
of hot intra-cluster media, such as temperature and chemical composition.
HST/ACS gives us deep images of high spatial resolution, and hence provides
essential morphological information of galaxies.
Our sample also includes a cluster at the NEP region where the visibility
of ASTRO-F (Japanese Space NIR--FIR mission, providing 10$'$$\times$10$'$
field-of-view; Shibai 2003; Murakami 2005)
is good; we plan for a follow-up observation to search for dusty
star-burst phenomena and to improve our photometric redshifts.
Also note that all of our clusters above $z=0.6$ are targets for
space-IR imaging by the Spitzer Telescope.
We also include seven clusters that have existing detailed
Sunyaev--Zeldovich effect (S--Z) maps for comparing of the
sub-structures in the cluster cores.
\item Visibility: The maximum elevation must be $>40^\circ$
(airmass $<$ 1.6) from the Subaru Telescope at Mauna Kea in Hawaii for
efficient observations.
\end{itemize}

Note that our total sample includes some GTO (guaranteed time observations)
targets of the Suprime-Cam team (CL~1604 and RX~J0849) and those targeted
by different teams, including some of our members (CL~0024, CL~0939, MS~0451,
and MS1054).
These are appended
to the PISCES targets, since the imaging data of these distant clusters
are taken under similar scientific motivations, and the data are
useful if viewed under the same context and in a systematic way.

\begin{table*}
\caption{
Total PISCES targets (planned).
}
\small
\begin{tabular}[h]{llllllllll}
\hline\hline
Class & Cluster          & \hspace{-0.3cm} \hspace{0.2cm} RA      & \hspace{0.2cm} Dec      & $z$ & L$_{\rm X}$ & Bands & Coordination & Status \\
($z$) &                  & \hspace{-0.3cm}  (J2000)    &   (J2000)    &  & 10$^{44}$ & & & &\\
\hline
0.4 & CL~0024+1652 & \hspace{-0.3cm}       00 26 35.7 & \hspace{-0.3cm} +17 09 43   & 0.39  &  3.2 & \hspace{-0.3cm} $BRz'$,NB & \hspace{-0.3cm} ACS, XMM, Cha & \hspace{-0.2cm} Kodama+ '04\\
& CL~0939+4713 & \hspace{-0.3cm}       09 42 56.6 & \hspace{-0.3cm} +46 59 22     & 0.41  &  9.2 & \hspace{-0.3cm} $BVRI$,NB & \hspace{-0.3cm} XMM & \hspace{-0.2cm} Kodama+ '01\\
& RX~J2228.5+2036 & \hspace{-0.3cm}      22 28 34.4   & \hspace{-0.3cm} +20 36 47     & 0.41  & 16.5 & \hspace{-0.3cm} $BVRi'$   & \hspace{-0.3cm} Cha, S-Z & \hspace{-0.2cm} Planned\\
\hline
0.55 & MS~0451.6$-$0305 & \hspace{-0.3cm}   04 54 10.9 & \hspace{-0.3cm} $-$03 01 07   & 0.55  & 12.0 & \hspace{-0.3cm} $BVRI$    & \hspace{-0.3cm} ACS, S-Z & \hspace{-0.2cm} Partly done\\
& CL~0016+1609 & \hspace{-0.3cm}       00 18 33.3 & \hspace{-0.3cm} +16 26 36   & 0.55 & 26.0$^\dagger$ & \hspace{-0.3cm} $BVRi'z'$ & \hspace{-0.3cm} ACS, XMM, Cha, S-Z & \hspace{-0.2cm} {\bf This paper}\\
& MS~2053.7$-$0449 & \hspace{-0.3cm}   20 56 22.4 & \hspace{-0.3cm} $-$04 37 43 & 0.58 &  5.0 & \hspace{-0.3cm} $BVRi'z'$ & \hspace{-0.3cm} ACS, XMM, S-Z & \hspace{-0.2cm} Partly done\\
\hline
0.85 & RX~J1716.4+6708 & \hspace{-0.3cm}    17 16 49.6 & \hspace{-0.3cm} +67 08 30     & 0.81 &  2.7$^\ddagger$ & \hspace{-0.3cm} $VRi'z'$  & \hspace{-0.3cm} Cha, ASTRO-F, Spi & \hspace{-0.2cm} Planned\\
& MS~1054.4$-$0321 & \hspace{-0.3cm}   10 57 00.2 & \hspace{-0.3cm} $-$03 37 27 & 0.82  & 20.0 & \hspace{-0.3cm} $VRi'z'$  & \hspace{-0.3cm} ACS, XMM, Cha, S-Z, Spi & \hspace{-0.2cm} Sato+ '03\\
& RX~J0152.7$-$1357 & \hspace{-0.3cm}  01 52 41.0 & \hspace{-0.3cm} $-$13 57 45 & 0.83  & 16.0 & \hspace{-0.3cm} $VRi'z'$  & \hspace{-0.3cm} ACS, XMM, Cha, S-Z, Spi & \hspace{-0.2cm} {\bf This paper}\\
& CL~J1226.9+3332 & \hspace{-0.3cm}    12 26 58.0 & \hspace{-0.3cm} +33 32 54     & 0.89   & 53.0 & \hspace{-0.3cm} $VRi'z'$  & \hspace{-0.3cm} ACS, XMM, Cha, S-Z, Spi & \hspace{-0.2cm} Planned\\
& CL~J1604+4321 & \hspace{-0.3cm}         16 04 31.5 & \hspace{-0.3cm} +43 21 17& 0.92  &  2.0 & \hspace{-0.3cm} $VRi'z'$  & \hspace{-0.3cm} ACS, XMM, Spi & \hspace{-0.2cm} Partly done\\
\hline
1.2 & RDCS~J0910+5422 & \hspace{-0.3cm}    09 10 00.0 & \hspace{-0.3cm} +54 22 00   & 1.11  &  2.1 & \hspace{-0.3cm} $VRi'z'$  & \hspace{-0.3cm} ACS, Cha, Spi & \hspace{-0.2cm} Planned\\
& RDCS~J1252$-$2927 & \hspace{-0.3cm}     12 52 54.4 & \hspace{-0.3cm} $-$29 27 17 & 1.24  &  6.6 & \hspace{-0.3cm} $VRi'z'$  & \hspace{-0.3cm} ACS, XMM, Cha, Spi & \hspace{-0.2cm} Planned\\
& RX~J1053.7+5735 & \hspace{-0.3cm}    10 53 39.8 & \hspace{-0.3cm} +57 35 18     & 1.14  &  2.0$^\ddagger$ & \hspace{-0.3cm} $VRi'z'$  & \hspace{-0.3cm} Cha, Spi & \hspace{-0.2cm} Planned\\
& RX~J0848.9+4452 & \hspace{-0.3cm}    08 48 56.3 & \hspace{-0.3cm} +44 52 16     & 1.26  &  2.8 & \hspace{-0.3cm} $BVRi'z'$ & \hspace{-0.3cm} ACS, XMM, Cha, Spi & \hspace{-0.2cm} Nakata+ '05\\
\hline
\end{tabular}\\
L$_{\rm X}$ shows the bolometric X-ray luminosity in units of
10$^{44}$erg s$^{-1}$ ($H_0=70$) with some exceptions
($^\dagger$ 0.4--10 kev, $^\ddagger$ 0.5--2 kev).
`NB' in the 7th column indicates narrow-band imaging to target H$\alpha$ line
at the cluster redshift.
In the 8th column,
`XMM' and `Cha' indicate the XMM and Chandra targets, respectively.
`ACS' indicates the HST/ACS targets and
`S-Z' indicates the Sunyaev-Zeldovich effect targets in Radio.
`ASTRO-F'
and `Spi' indicate the targets for ASTRO-F and Spitzer space-IR
telescopes, respectively.\\
Refs.\ for $L_X$: CL~0024 (Ota et al.\ 2003); CL~0939
(De Filippis et al.\ 2003);
RX~J2228 (Pointecouteau et al.\ 2002), MS~0451 (Donahue \& Stocke 1995);
CL~0016 (Worrall \& Birkinshaw 2003); MS~2054 (Henry 2000);
RX~J1716 (Henry et al.\ 1997);
MS~1054 (Jeltema et al.\ 2001); RX~J0153 (Maughan et al.\ 2003);
RX~J1227 (Maughan et al.\ 2004); CL~1604 (Lubin et al.\ 2004);
RDCS~J0910 (Stanford et al.\ 2002); CL~1252 (Rosati et al.\ 2004);
RX~J1054 (Hasinger et al.\ 1998); and RX~J0849 (Stanford et al.\ 2001).\\
Refs.\ for S-Z: MS~1054, RX~J0153, RX~J1227 (Joy et al.\ 2001); MS~0451,
CL~0016, MS~2054, MS~1054 (Grego et al.\ 2001);
RX~J2228 (Pointecouteau et al.\ 2002).
\label{tab:targets}
\end{table*}

\subsection{This Paper}

In 2003 September, as a part of the PISCES project, we successfully obtained
complete imaging data-set for the two clusters CL~0016+1609
($z=0.55$; hereafter CL~0016) and RX~J0152.7$-$1357
($z=0.83$; hereafter RX~J0153)
In this paper, we concentrate on three clusters in total;
the above newly observed two clusters CL~0016, RX~J0153, and a previous target
CL~0939+4713 ($z=0.41$; hereafter CL~0939) for comparison
(Kodama et al.\ 2001).
These three clusters under investigation are located in a sequence of
redshifts corresponding to lookback times of 4.3, 5.4, and 7.0 Gyrs,
and are therefore suited for us to look for possible signatures of
evolutionary effects in clusters.

The first results on the CL~0939+4713 cluster were already reported by
Kodama et al.\ (2001), showing large-scale filamentary structures around
this cluster and a transition of galaxy
colors in the group environment along the filaments.
Note that Iye et al.\ (2000) also presented their Subaru/Suprime-Cam
image of CL~0939+4713 taken in their GT phase.
This paper expands Kodama et al.'s (2001) analysis (for
CL~0939+4713) to two additional higher redshift clusters, concentrating
on the large-scale structures.
The environmental dependence of the photometric properties of galaxies in
these two clusters and those in local counterparts from SDSS
(Sloan Digital Sky Survey; York et al. 2000) will be presented
in Tanaka et al.\ (2005).

We adopt the cosmological parameters ($h_{70}$, $\Omega_m$,
$\Omega_{\Lambda}$)=(1.0, 0.3, 0.7) throughout this paper, where
$h_{70}$ is defined as $H_0$/(70~km s$^{-1}$Mpc$^{-1}$).
With this parameter set, 1$'$ corresponds to physical scales of
0.33, 0.38, and 0.46~Mpc, or comoving scales of
0.46, 0.59, and 0.84~Mpc, at cluster redshifts of 0.41, 0.55, and 0.83,
respectively.
All of the magnitudes used in this paper will be given in the
AB-magnitude system.

\section{Observations, Reduction, and Analysis}
\label{sec:obs}

\subsection{Observations and Reduction}

\begin{table*}
\caption{Summary of the observations.$^\ast$}
\begin{center}
\begin{tabular}{lccccc}
\hline\hline
Cluster & Passband & Date & Exp. times & Lim. mag.   & Seeing \\
        &          &      & (total)    & ($5\sigma$) & (FWHM) \\
\hline
CL~0939+4713 & $B$  & 2001 Jan 21--22 & 60 min & 26.6 & $\sim1.$\hspace{-2pt}$''$1 \\
            & $V$  & 2001 Jan 21--22 & 36 min & 26.0 & $\sim0.$\hspace{-2pt}$''$7 \\
            & $R$  & 2001 Jan 21--22 & 66 min & 26.0 & $\sim1.$\hspace{-2pt}$''$0 \\
            & $I$  & 2001 Jan 21--22 & 21 min & 25.5 & $\sim0.$\hspace{-2pt}$''$7 \\
\hline
CL~0016+1609 & $B$  & 2003 Sep 25--26 & 90 min & 26.9 & $\sim0.$\hspace{-2pt}$''$65 \\
            & $V$  & 2003 Sep 25--26 & 96 min & 26.2 & $\sim0.$\hspace{-2pt}$''$65 \\
            & $R$  & 2003 Sep 25--26 & 64 min & 26.0 & $\sim0.$\hspace{-2pt}$''$65 \\
            & $i'$ & 2003 Sep 25--26 & 60 min & 25.9 & $\sim0.$\hspace{-2pt}$''$65 \\
            & $z'$ & 2003 Sep 25--26 & 47.5 min & 24.6 & $\sim0.$\hspace{-2pt}$''$65 \\
\hline
RX~J0152.7$-$1357 & $V$ & 2003 Sep 25--26, 2003 & 120 min & 26.7 & $\sim0.$\hspace{-2pt}$''$65 \\
            & $R$  & 2003 Sep 25--26, 2003 & 116 min & 26.5 & $\sim0.$\hspace{-2pt}$''$65 \\
            & $i'$ & 2003 Sep 25--26, 2003 & 75 min & 26.1 & $\sim0.$\hspace{-2pt}$''$65 \\
            & $z'$ & 2003 Sep 25--26, 2003 & 77 min & 25.0 & $\sim0.$\hspace{-2pt}$''$65 \\
\hline
\end{tabular}\\
\small
$^\ast$ Limiting magnitudes are measured within 2$''$ apertures.
Seeing sizes are measured from the combined frames.
\end{center}
\label{tab:observation}
\end{table*}

Imaging data of CL~0016 and RX~J0153 were obtained with
Suprime-Cam (0.\hspace{-2pt}$''$202 per pixel, and
34\arcmin\ $\times$ 27\arcmin\
field of view) mounted on the Subaru Telescope at Mauna Kea during
the nights of 2003 September 25--26.
Images were obtained through several broad-band filters
($BVR i^\prime z^\prime$ and $VR i^\prime z^\prime$, respectively).
The net exposure times and the 5-$\sigma$ limiting magnitudes
are summarized in table~\ref{tab:observation}.
The seeing was extremely stable during these nights and across
the bands, and its sizes were always between
0.\hspace{-2pt}$''$5 and 0.\hspace{-2pt}$''$65 (FWHM).
Therefore, our combined frames achieved the uniform seeing size of
0.\hspace{-2pt}$''$65 in all passbands for both clusters.
The sky conditions were photometric, and the photometric zero-points were
calibrated based on the Landolt (1992) standard stars in $B$, $V$,
and $R$-bands,
and $i'$ and $z'$-band images were calibrated directly onto the
SDSS system (Fukugita et al.\ 1995)
based on the spectro-photmetric standard stars.
After correcting for the Galactic extinction (Schlegel et al.\ 1998),
we checked the accuracy of our photometric zero-points.
We compared the colors of Galactic stars in our field with
those of Gunn and Stryker (1983)
stars (stellar SEDs convolved with Suprime-Cam responses).
It turned out that our colors are slightly shifted with respect to the GS stars
($\sim0.1$ magnitude at most).
Although we do not know the exact reason for this discrepancy,
it may be partly because the filter response functions, including
CCD sensitivity curves, are slightly different between SDSS and our
observations with Suprime-Cam.
We shifted the zero-points of the photometric catalogues so as to match
with the GS star colors in the same manner as in Ajiki et al.\ (2003).

\begin{figure*}
\begin{center}
\vspace{2cm}
{\large\bf 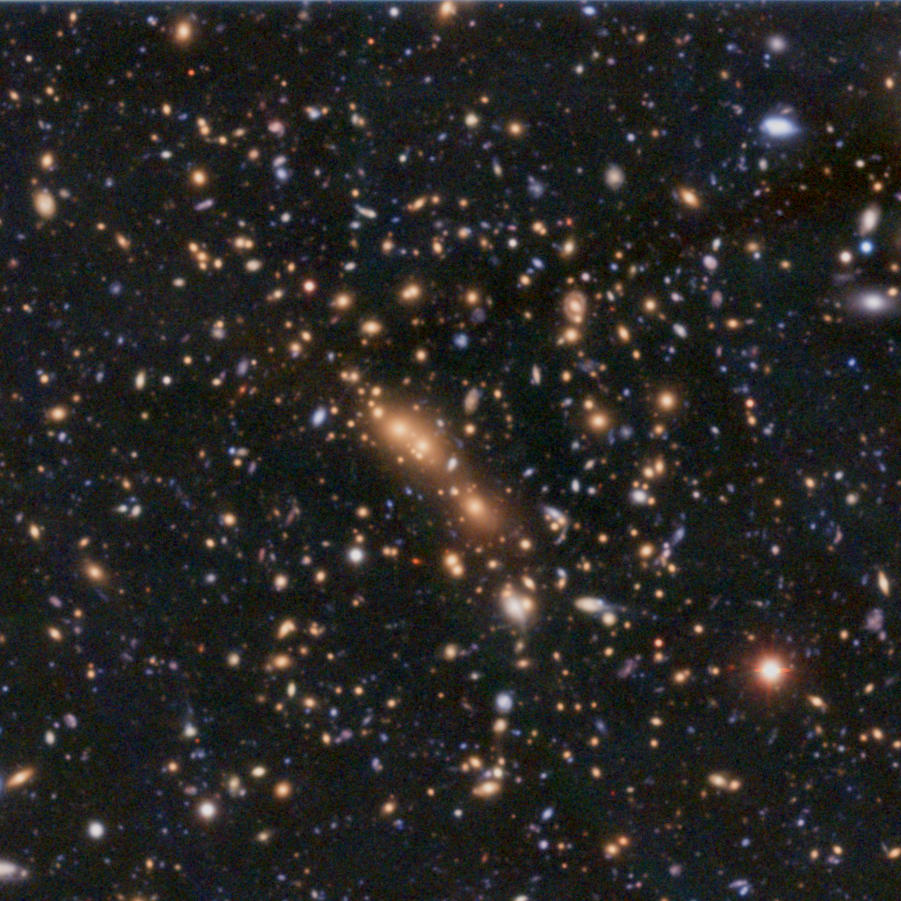}
\vspace{2cm}
\end{center}
\caption{
False-color image of the central 3$'$ $\times$ 3$'$ region of
CL~0016+1609 constructed from our $V$, $R$, and $i'$ images.
North is up and east is to the left.
}
\label{fig:cl0016_image}
\end{figure*}

\begin{figure*}
\begin{center}
\vspace{2cm}
{\large\bf 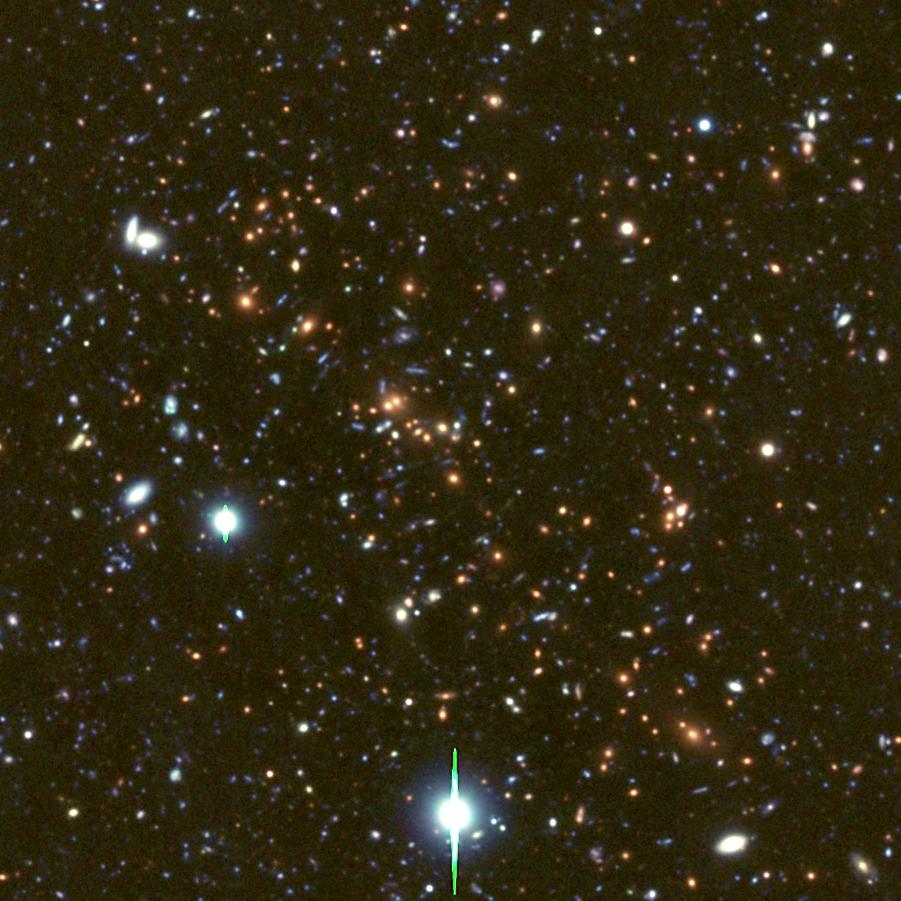}
\vspace{2cm}
\end{center}
\caption{
False-color image of the central 3$'$ $\times$ 3$'$ region of
RX~J0152.7$-$1357 constructed from our $V$, $R$, and $i'$ images.
North is up and east is to the left.
}
\label{fig:rxj0153_image}
\end{figure*}

The data were reduced with the {\sc IRAF} and {\sc NEKOSOFT}
(Yagi et al.\ 2002) software packages, following standard procedures of bias
subtraction and flat-fielding.  The latter was achieved
using supersky (self) flats constructed from the median of the dithered science
frames (objects are masked).
We then combined the mosaiced chips, carefully matching
the relative flux among them.
Note that, since the seeing was very stable during the nights,
we performed no point-spread function (PSF) equalization in order to
avoid image degradation.
Color pictures of the central regions (3\arcmin $\times$ 3\arcmin)
of the two clusters are shown in figures~\ref{fig:cl0016_image} and
\ref{fig:rxj0153_image}.

Galaxies are detected in the $i^\prime$ and $z^\prime$-band images for
CL~0016 and RX~J0153, respectively, using the {\sc SExtractor}
software (Bertin and Arnouts 1996).
We detected all objects with at least 8 connected pixels
(0.33 arcsec$^2$, equivalent to the area of the PSF) 
more than 2 $\sigma$ above the median sky.
The images in all filters were aligned by centroiding
stars throughout the field and fitting a stretch and shift.
We used a fixed 2\arcsec\ diameter aperture
(corresponding to 13--15~kpc at the cluster redshifts) when measuring the
galaxy colors.  The magnitude {\sc MAG\_AUTO} was used as a measure
of the total magnitude.

\subsection{Photometric Redshifts}
\label{sec:photz}

In order to map the structures of clusters out to low-density regions
around the cluster cores, the key requirement is the removal of
unassociated galaxies in the foreground and background that dominate
at the outskirts of clusters.
This is important to maximize the contrast of the member
galaxies on the projected sky.

Spectroscopy is, of course, the ideal method to remove field contamination
from our sample.  It is not practical, however, to obtain spectroscopy for
$\gg$ 10000 very faint galaxies required for our analysis.
We therefore exploit the photometric redshift (phot-$z$) technique,
as an observationally
efficient and reliable method to largely remove contamination
($\sim$ 90\% over the whole Suprime-Cam field).
We applied the photometric redshift code developed by Kodama, Bell, and Bower
(1999) to the galaxies in our full photometric catalog in $BVRI$
(CL~0939), $BVRi'z'$ (CL~0016), and $VRi'z'$ (RX~J0153), respectively.
We note that these combinations of passbands neatly cover the 4000~\AA\, break
features at the rest-frame of each cluster, which is essential to obtain
accurate photometric redshifts, especially for red cluster members
(Kodama et al.\ 1999); typical uncertainties of only
$|\Delta z|=$ 0.05--0.1 are expected.
The photometric redshift technique is now widely used to trace large-scale
structures in the distant Universe (e.g., Connolly et al. 1996;
Kodama et al. 2001; Gray et al. 2003; Nakata et al. 2005).

In order to apply the photometric redshift technique,
the colors of the model galaxies (templates)
and those of the observed galaxies must match perfectly well.
In reality, a zero-point mismatch between the observed data and the model
spectra often exists.
First of all, an absolute flux calibration in the observed data
is difficult with only a small number of photometric standards
observed during the nights (1--2 per band).
Secondly, the synthesized model spectra inevitably have uncertainties
on the order of 0.05--0.2 magnitudes (Charlot et al.\ 1996).
Taking these uncertainties into account,
we shift our model magnitudes
by 0.00--0.15 magnitudes so that the models can reasonably reproduce the
observed colors of the red color--magnitude sequences in the cluster
cores, because the red-sequence colors of clusters are known to be
good guides of passive evolution (e.g., Kodama et al.\ 1998).

The definition of our photometric members (phot-$z$ members) for CL~0939 is
$0.32\le z\le 0.48$ or $-0.09\le\Delta z\le 0.07$, as used in
Kodama et al.\ (2001). However, for the other two clusters
(CL~0016 and RX~J0153), we narrowed the redshift ranges to
$0.50\le z\le 0.58$ and $0.78\le z\le 0.86$, respectively, or
$-0.05\le\Delta z\le 0.03$, to further suppress any remaining
contamination and to achieve higher contrasts on true structures in the
lower density regions located at the cluster redshifts.
The depths of the phot-$z$ sliced regions are 553, 254, and 215 Mpc
in the comoving scale for CL~0939, CL~0016, and RX~J0153, respectively.
It should be noted that these stringent criteria for photometric
members tend to lose some true cluster members due to the error in
photometric redshifts, which can be larger than $|\Delta z|>0.05$
especially for blue members.
When we analyze the galaxy properties in a future paper (Tanaka et al.\ 2005),
we will adopt much broader ranges for photometric
members to achieve higher completeness at the cost of higher contamination
(the remaining contamination will, however, be subtracted statistically
as in Kodama et al.\ 2001).
In the current work, we did not subtract the remaining contamination,
since we concentrated on the structures of the galaxy distribution, and 
did not primarily consider the photometric properties of the member galaxies.

A comparison of our photometric redshifts and the spectroscopic redshifts
available from the literature are briefly discussed in Kodama et al. (2001)
for CL~0939, and will be reported in Tanaka et al. (2005) for CL~0016 and
RX~J0153.
In short, we found that about 70--80\% of the spectroscopic members could be
assigned reasonable photometric redshifts, and were hence identified as
our photometric members.
We also note that a spectroscopic follow-up of the CL~0016 and RX~J0153 clusters
with FOCAS on Subaru was performed in 2004 October and the data are being
processed, which will confirm membership for the bright objects.

\section{Results}
\label{sec:results}

We first show color--magnitude diagrams in figure~\ref{fig:cmr}
to demonstrate the data that we have obtained in all available passbands.
We then use these diagrams to separate the red and blue cluster members.
The 2-D distributions of these photometric members are shown on two scales,
close-up of the central regions of the clusters (figure~\ref{fig:map_center})
to trace the sub-structures in or near the cluster cores
and the whole Suprime-Cam field ($\sim$30\arcmin\ across;
figure~\ref{fig:map_lss}) to map out the large-scale structures surrounding
the cores.
The optical sub-structures of the
clusters are compared to the X-ray maps and show
similarities (figure~\ref{fig:map_center}).
We will then try to quantify the structures of the clusters using the
shapes of the 2-D iso-density contours (figure~\ref{fig:contour_shape})
and a Fourier expansion
of the galaxy distribution in the tangential direction around the
cluster cores (figure~\ref{fig:fourier})
to discuss the roundness and filamentarity of the structures.
Finally, these observed structures will be compared to the model predictions
from our numerical simulation ($\nu$GC; Nagashima et al.\ 2005)
which combines the $N$-body calculation concerning the
development of the dark matter halos and what is commonly known as
semi-analytic modelling of galaxy formation and evolution
(star formation and mergers) within the halos
(figures~\ref{fig:map_lss} and \ref{fig:contour_shape}).

\subsection{Color--Magnitude Diagrams}
\label{sec:cmd}

\begin{figure*}
\begin{center}
\vspace{2cm}
{\large\bf 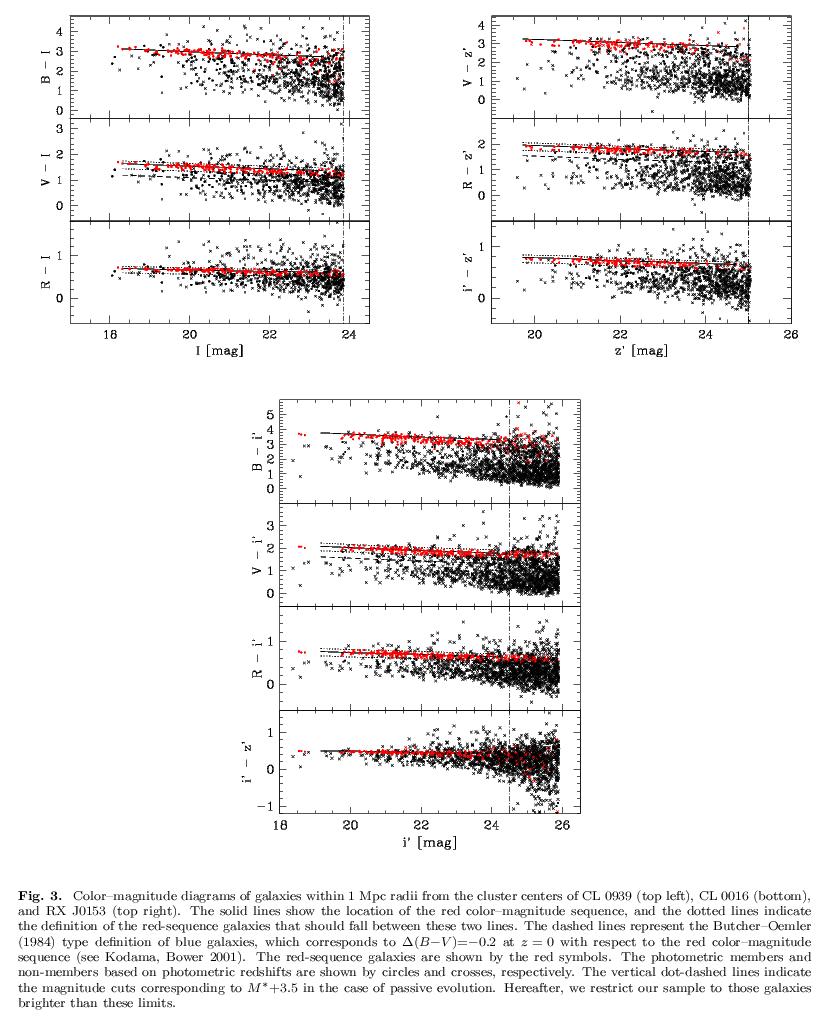}
\vspace{2cm}
\end{center}
\caption{
Color--magnitude diagrams of galaxies within 1~Mpc radii from the
cluster centers of CL~0939 (top left), CL~0016 (bottom),
and RX~J0153 (top right).
The solid lines show the location of the red color--magnitude sequence,
and the dotted lines indicate the definition of the red-sequence galaxies
that should fall between these two lines.
The dashed lines represent the Butcher--Oemler (1984) type definition of blue
galaxies, which corresponds to $\Delta$($B$$-$$V$)=$-$0.2 at $z=0$ with respect
to the red color--magnitude sequence (see Kodama, Bower 2001).
The red-sequence galaxies are shown by the red symbols. The photometric
members and non-members based on photometric redshifts are shown by circles
and crosses, respectively.
The vertical dot-dashed lines 
indicate the magnitude cuts corresponding to $M^*$+3.5 in the case of
passive evolution.  Hereafter, we restrict our sample to those galaxies
brighter than these limits.
}
\label{fig:cmr}
\end{figure*}

Color--magnitude diagrams of the galaxies within 1~Mpc radii 
from the centers of the three clusters are shown
in figure~\ref{fig:cmr}.
The photometric members selected based on the photometric redshifts
are indicated by circles, while the photometric non-members are indicated by
crosses.  As expected, the photometric members show a tight color--magnitude
relations in all colors that are composed of old passively evolving
galaxies (e.g., Bower et al.\ 1992; Ellis et al.\ 1997;
Stanford et al.\ 1998; Kodama et al.\ 1998; Terlevich et al.\ 2000).
The solid lines show the predicted color--magnitude relations at the cluster
redshifts from Kodama et al.\ (1998), which are constructed so as to reproduce
the color--magnitude relation of Coma elliptical galaxies
(Bower et al.\ 1992).
The zero-points of the models are corrected as in subsection 2.2.
We use these diagrams to define `red' galaxies, `blue' galaxies, and
`red-sequence' galaxies, based on the distance from the red color--magnitude
sequence (solid lines).
The galaxies bracketed by the dotted lines are defined
to be `red-sequence' galaxies, which are plotted in the middle-row panels
of figure~\ref{fig:map_lss}.
The exact definitions of the red-sequence galaxies are given by the
following two color ranges from the red-sequence for each cluster:

\noindent
{\bf CL~0939:}
\begin{equation}
\hspace{-0.5cm} -0.20 < (V-I) - [1.74-0.0547\times (I-18.32)] < 0.10
\end{equation}
and
\begin{equation}
\hspace{-0.5cm} -0.10 < (R-I) - [0.62-0.0174\times (I-18.32)] < 0.05.
\end{equation}
{\bf CL~0016:}
\begin{equation}
\hspace{-0.5cm} -0.20 < (V-i') - [2.09-0.0627\times (i'-19.16)] < 0.14
\end{equation}
and
\begin{equation}
\hspace{-0.5cm} -0.10 < (R-i') - [0.76-0.0251\times (i'-19.16)] < 0.07.
\end{equation}
{\bf RX~J0153:}
\begin{equation}
\hspace{-0.5cm} -0.20 < (R-z') - [1.96-0.0575\times (z'-19.72)] < 0.10
\end{equation}
and
\begin{equation}
\hspace{-0.5cm} -0.10 < (i'-z') - [0.79-0.0275\times (z'-19.72)] < 0.05.
\end{equation}

Although red-sequence galaxies are defined in the above two colors each,
they also show consistently red colors and form clear sequences in the other
colors as well.
The `red' galaxies and `blue' galaxies are separated by the dashed line
on the color--magnitude diagram which shows the color bracketing the
rest-frame 4000\AA\, break.
The distance of the separative line from the red color--magnitude
sequence corresponds to $\Delta$($B$$-$$V$)=$-$0.2 in the rest-frame
following the original Butcher and Oemler's (1984) type definition.
This is transformed to each observed frame using
Kodama et al.'s (1998) population synthesis model by taking into account
the color evolution (Kodama, Bower 2001).
The thus-defined red and blue galaxies are shown by different symbols in
figures~\ref{fig:map_center} and \ref{fig:map_lss}.

The vertical dot-dashed lines correspond to $M^*$+3.5 at the cluster
redshifts assuming passive evolution; hereafter, we restrict
our sample to those galaxies brighter than these limits.

\subsection{Substructures in the Cluster Cores}
\label{sec:sub}

\begin{figure*}
\begin{center}
\vspace{2cm}
{\large\bf 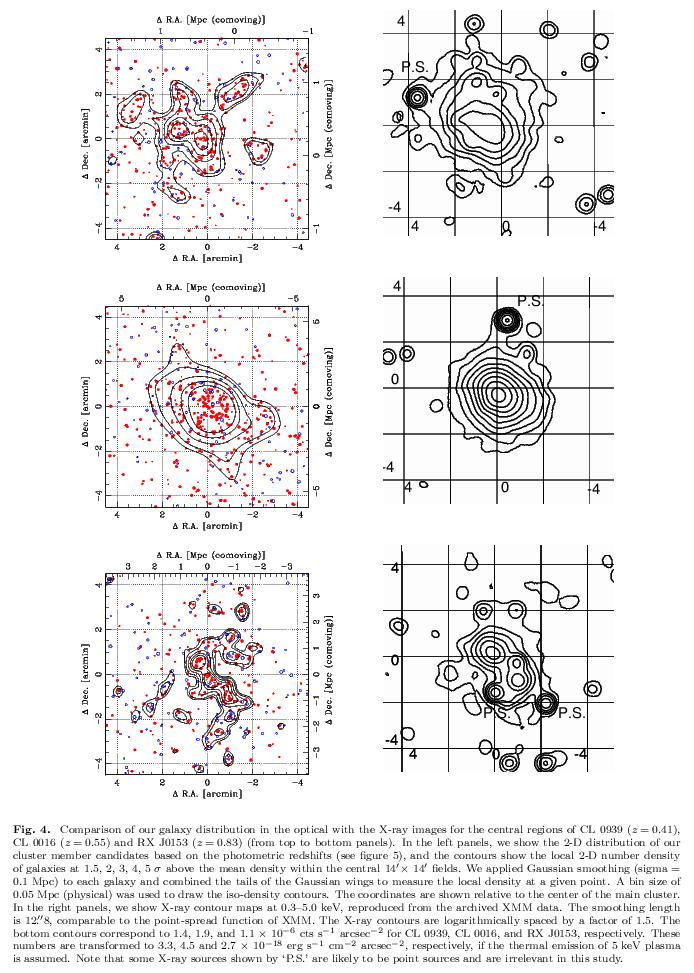}
\vspace{2cm}
\end{center}
\caption{
Comparison of our galaxy distribution in the optical with the X-ray images
for the central regions of
CL~0939 ($z=0.41$), CL~0016 ($z=0.55$) and RX~J0153 ($z=0.83$)
(from top to bottom panels).
In the left panels, we show the 2-D distribution of our cluster member
candidates based on the photometric redshifts (see figure~\ref{fig:map_lss}),
and the contours show the local 2-D number density of
galaxies at 1.5, 2, 3, 4, 5 $\sigma$ above the mean density within the
central 14\arcmin $\times$ 14\arcmin\ fields.
We applied Gaussian smoothing (sigma = 0.1~Mpc) to each galaxy and combined
the tails of the Gaussian wings to measure the local density at a given point.
A bin size of 0.05~Mpc (physical) was used to draw the iso-density contours.
The coordinates are shown relative to the center of the main cluster.
In the right panels, we show X-ray contour maps at 0.3--5.0 keV,
reproduced from the archived XMM data.
The smoothing length is 12.\hspace{-2pt}$''$8, comparable to the point-spread
function of XMM. 
The X-ray contours are logarithmically spaced by a factor of 1.5.
The bottom contours correspond to 1.4, 1.9, and 1.1
$\times$ 10$^{-6}$ cts s$^{-1}$ arcsec$^{-2}$ for CL~0939, CL~0016,
and RX~J0153, respectively.  These numbers are transformed to
3.3, 4.5 and 2.7 $\times$ 10$^{-18}$ erg s$^{-1}$ cm$^{-2}$ arcsec$^{-2}$,
respectively, if the thermal emission of 5~keV plasma is assumed.
Note that some X-ray sources shown by `P.S.' are likely to be point sources
and are irrelevant in this study.
}
\label{fig:map_center}
\end{figure*}

The 2-D distribution of the photometric members defined as in
subsection \ref{sec:photz} in the central regions of the clusters, are shown
in the left panels of figure~\ref{fig:map_center}.
The substructures are clearly seen in all three clusters under study.
We use the archived XMM data of these three clusters and reproduce the
X-ray contour maps at 0.3--5.0 keV in the right panels for a comparison.
These data were originally presented
in De Filippis, Schindler and Castillo-Morales (2003),
Worral and Birkinshaw (2003), and Jones et al.\ (2004)
for CL~0939, CL~0016 and RX~J0153, respectively.
The similarity of the structures in the cluster cores and their neighbouring
regions between the optical galaxy distribution and the current X-ray images
(hot gas trapped in the potential well) is seen as described
individually, as follows:

\begin{itemize}
\item {\bf CL~0939:}
The core structure is composed of two major clumps stretching from
East to West and to SW.
Many filaments to North, NW, South and NE are also clearly identified
in the optical image. In the X-ray images only the North and the NW
extensions are visible.
We also note that the similarity is also seen between the weak lensing mass
map and the {\it Rosat} X-ray image (Iye et al.\ 2000).
\item {\bf CL~0016:}
The galaxy distribution is rather round and featureless within the
central $\sim$2\arcmin\ radius region.  Beyond this area the optical
distribution tends to be stretched in the NE--SW direction which links to the
distinct SW clump at ($\Delta$R.A., $\Delta$Dec.)=(4$'$,$-$8.5$'$)
(\S3.3; figure\ 5).
\item {\bf RX~J0153:}
The central three red clumps are aligned linearly in the NE--SW
direction forming a chain like structure.  This filament corresponds to
two major clumps in the X-ray image.  Also, the NW--SE filament
perpendicular to the NE--SW filament crossing at the NE edge is 
seen both in the optical and X-ray images.
It is notable that the southern extension of the filament is seen in
the optical image, but is not clear in the X-ray image.
The small clump at ($\Delta$R.A., $\Delta$Dec.) = (4$'$,$-$0.\hspace{-2pt}$'$9)
has been recently confirmed to lie at the cluster redshift by Demarco et al.\ (2005).
\end{itemize}

\subsection{Large-Scale Structures beyond the Cores}
\label{sec:lss}

\begin{figure*}
\begin{center}
\leavevmode
\vspace{2cm}
{\large\bf 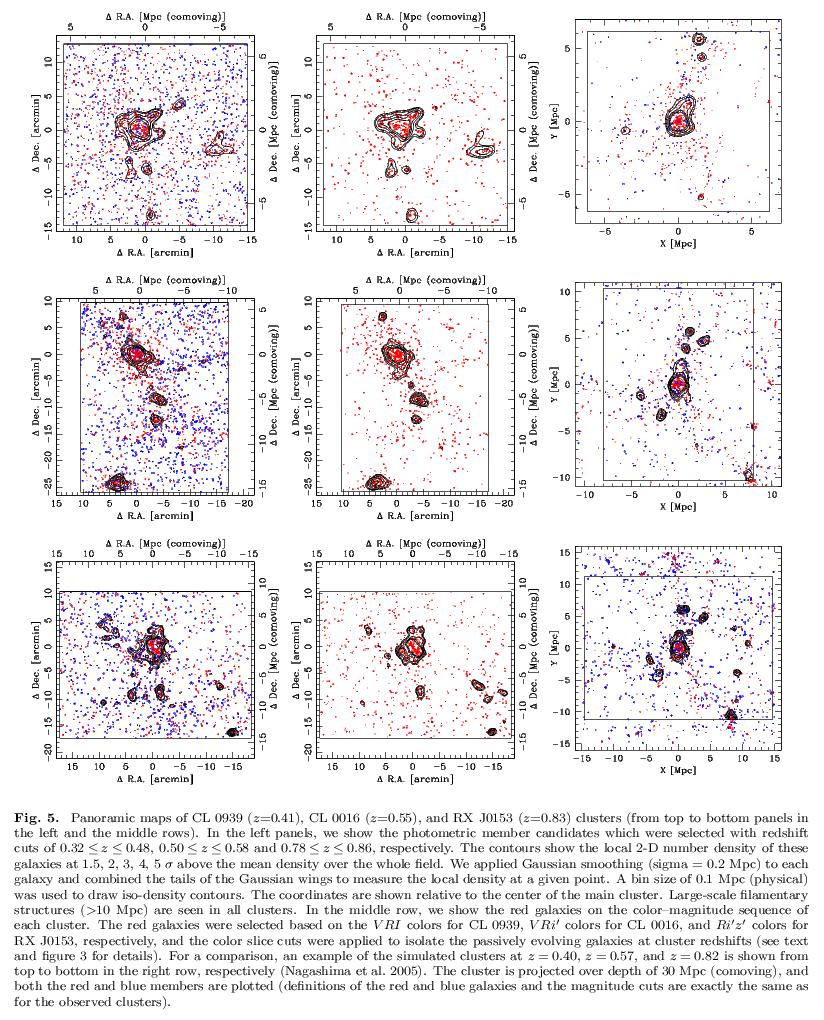}
\vspace{2cm}
\end{center}
\caption{
Panoramic maps of CL~0939 ($z$=0.41), CL~0016 ($z$=0.55),
and RX~J0153 ($z$=0.83) clusters (from top to bottom panels in the left
and the middle rows).
In the left panels, we show the photometric member candidates which were
selected with redshift cuts of $0.32\le z\le 0.48$,
$0.50\le z\le 0.58$ and $0.78\le z\le 0.86$, respectively.
The contours show the local 2-D number density of these galaxies at
1.5, 2, 3, 4, 5 $\sigma$ above the mean density over the whole field.
We applied Gaussian smoothing (sigma = 0.2~Mpc) to each galaxy and combined
the tails of the Gaussian wings to measure the local density at a given point.
A bin size of 0.1~Mpc (physical) was used to draw iso-density contours.
The coordinates are shown relative to the center of the main cluster.
Large-scale filamentary structures ($>$10~Mpc) are seen in all clusters.
In the middle row, we show the red galaxies on the color--magnitude
sequence of each cluster.  The red galaxies were selected based on the
$VRI$ colors for CL~0939, $VRi'$ colors for CL~0016,
and $Ri'z'$ colors for RX~J0153, respectively,
and the color slice cuts were applied to isolate the passively evolving
galaxies at cluster redshifts (see text and figure~\ref{fig:cmr} for details).
For a comparison, an example of the simulated clusters at $z=0.40$, $z=0.57$,
and $z=0.82$ is shown from top to bottom in the right row, respectively
(Nagashima et al.\ 2005).  The cluster is projected over depth of 
30~Mpc (comoving), and both the red and blue members are plotted (definitions of
the red and blue galaxies and the magnitude cuts are exactly the same as for
the observed clusters).
}
\label{fig:map_lss}
\end{figure*}

The galaxy distribution over the entire 30$'$ fields of the
three clusters is shown in figure~\ref{fig:map_lss}.
The photometric members are shown in the left panels separated,
into red and blue galaxies.
Another method that picks out the red-sequence galaxies has a
comparable sensitivity to the redshift ($\Delta z\sim 0.05$) for
old passively evolving populations at cluster redshift,
and hence it is a simple, but powerful, technique for finding structures,
such as clusters
(e.g., Gladders, Yee 2000; Lubin et al., 2000; Tanaka et al. 2001;
Ebeling et al.\ 2004).
The red-sequence galaxies defined by inequalities (1)--(6) are shown in the
middle panels.
An example of our model cluster (the most massive halo in our simulation)
is shown in the right panels for a comparison (see subsection 3.5).

The large-scale filamentary structures are clearly visible in all
clusters extending out
to the full Suprime-Cam fields, which correspond to $\sim$ 15--30~Mpc across
on the comoving scale.  Most of the structures found in the present study
are robust because they are both seen in the phot-$z$ members and in the red
sequence galaxies.
Also, the comoving volume within the photometric redshift slice over the
observed area is 0.84, 0.86, and 1.4 $\times$ 10$^5$ Mpc$^3$, for CL~0939,
CL~0016 and RX~J0153, respectively.
Given the fact that
the typical comoving density of clusters/groups ($>10^{14}M_{\odot}$) is
roughly 10$^{-5}$Mpc$^{-3}$ (Nagashima et al.\ 2005), the expected number of
clusters/groups that fall randomly within our narrow redshift slice would
be around unity per field.
Therefore, from a statistical point of view, most of the structures seen
in the redshift-sliced maps (figure~\ref{fig:map_lss}) are likely to be
in common large structures around the main bodies of the clusters, rather than
randomly falling into our phot-$z$ slices and being physically independent.

Since XMM provides a 30$'$ diameter field of view,
the XMM archived data enabled us to search for any diffuse X-ray emission
from the groups along the large-scale structures.
However, we did not find any significant emission from the groups outside the
3$'$ radii from the cluster centers, except for a known group,
RX~J0018.3+1618, in CL~0016 and the south group in RX~J0153 (see below).
The upper limits of the X-ray luminosities of these optically identified,
but X-rays undetected, groups were estimated to be 2--4, 1--5, and 10--20
in units of 10$^{43}$ erg s$^{-1}$ (0.5--10 keV) for CL~0939, CL~0016, and RX~J0153,
respectively.
Please note that some groups are out of the XMM field of view, or
very close to the edge, where sensitivity drops significantly.

We discuss below the large-scale structures of each cluster individually.
\begin{itemize}
\item {\bf CL~0939:}
As was reported in Kodama et al.\ (2001), although this cluster is dominated
by a red core, many filaments coming out from the core and extending
to large radii have also been identified.
This gives us an impression that the cluster is indeed located at the
node of the cosmic web, and that the surrounding material is being assembled
to the cluster core along these filaments, as the numerical simulations
suggest (e.g., Ghigna et al.\ 1998; Yahagi et al.\ 2004).
\item {\bf CL~0016:}
This field is known as a triple cluster in X-rays (ROSAT) and optical observations
(Connolly et al.\ 1996). The clumps at ($\Delta$R.A., $\Delta$Dec.) =
(0$'$,0$'$), ($-$4$'$,$-$8.\hspace{-2pt}$'$5) and (3.\hspace{-2pt}$'$5,$-$24$'$)
are all X-rays detected, and their physical associations have been confirmed
spectroscopically.
The central clump and the south clump are named RX~J0018.3+1618
(Hughes et al.\ 1995) and RX~J0018.8+1602 (Connolly et al.\ 1996), respectively.
We now find the structures connecting these three major clumps,
composed of several new clumps, such as that at ($-$3$'$, $-$12.\hspace{-2pt}$'$5),
and the large NE--SW filaments coming out from the richest cluster core.
All of these structures seem to align and form a huge connected filament
extending more than 20~Mpc (comoving).
\item {\bf RX~J0153:}
This cluster has a core of very complicated shape, and multiple filaments
have been identified coming out from the core, which may be connected to the
small clumps scattered around the core.
The SE clump at ($\Delta$R.A., $\Delta$Dec.) = (4$'$,$-$9$'$) was detected
in an XMM image, but it is likely to be a foreground structure at slightly
lower redshift, since the location of the red color sequence is slightly
offset to bluer color [by $\Delta (R-z')$=0.1~mag]
in this system compared to that of the main cluster.
In fact, this clump disappears in the middle panel of figure~\ref{fig:map_lss}
where only red-sequence galaxies consistent with the cluster redshift
are plotted.
In any case, this rather complicated structure in and around the core
(figures~\ref{fig:map_center} and \ref{fig:map_lss}) suggests that it is
still in a dynamically young stage of cluster-scale assembly at this high
redshift. This is in qualitative agreement with the build-up of clusters
in the CDM simulations (e.g., Ghigna et al.\ 1998; Yahagi et al.\ 2004).
\end{itemize}

\subsection{Morphology of Clusters: Quantifying the Structures}
\label{sec:morph}

To trace the evolution of cluster structures with time, and to compare
it with numerical simulations, we tried to quantify the structures of the
phot-$z$ member distribution by defining the following two quantities:
(1) roundness of the iso-density contour as a function of the
threshold of the contour;
(2) filamentarity of galaxy distribution in the tangential direction.
We intuitively expect that at high redshifts, clusters are in the process of
vigorous assembly by capturing the surrounding systems along the filaments,
and thus having more and more clumpy and filamentary structures in and around
the cluster cores and large deviation from a spherical distribution.

\begin{figure*}
\begin{center}
  \FigureFile(80mm,80mm){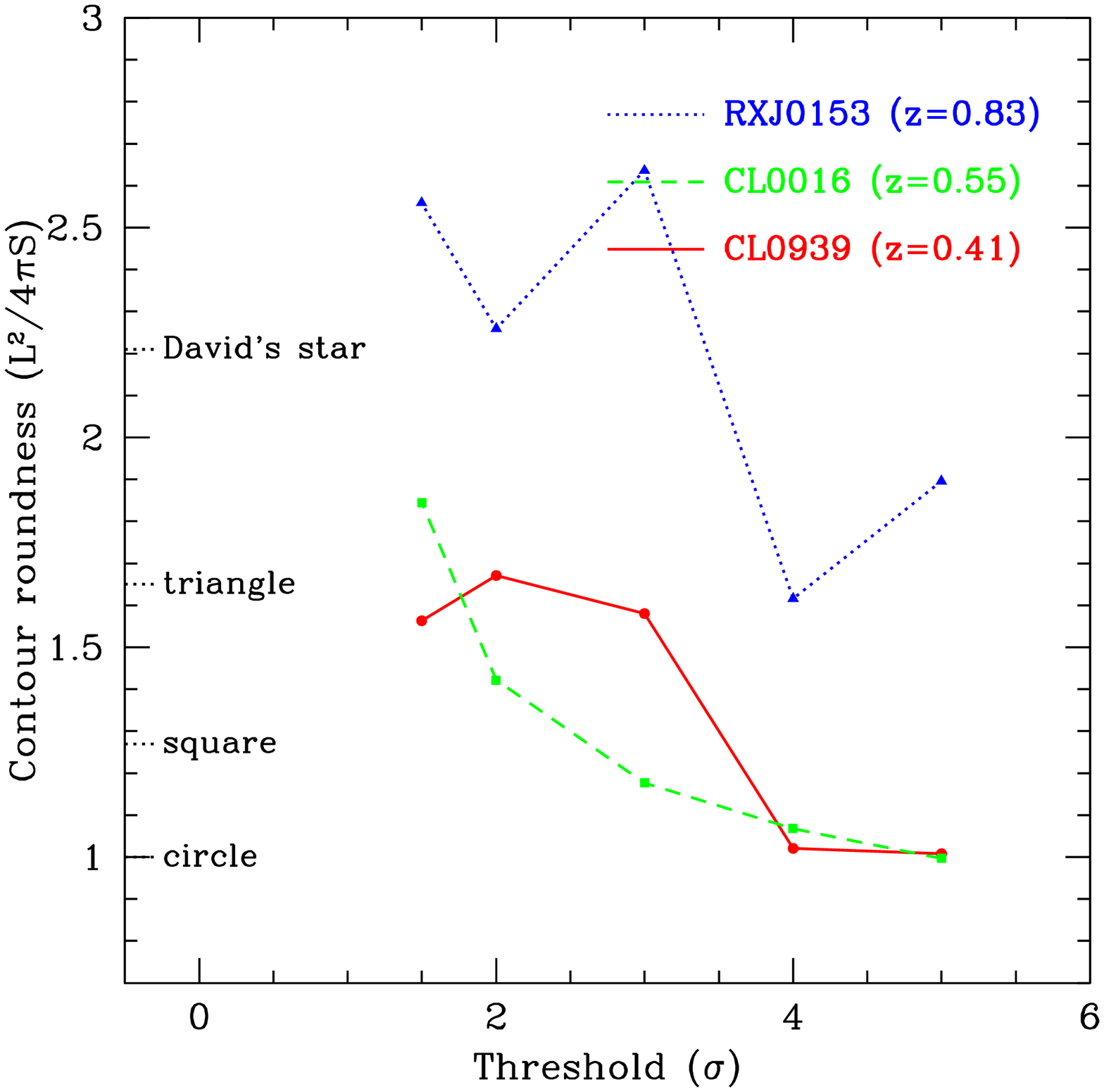}
  \hskip 0.5cm
  \FigureFile(80mm,80mm){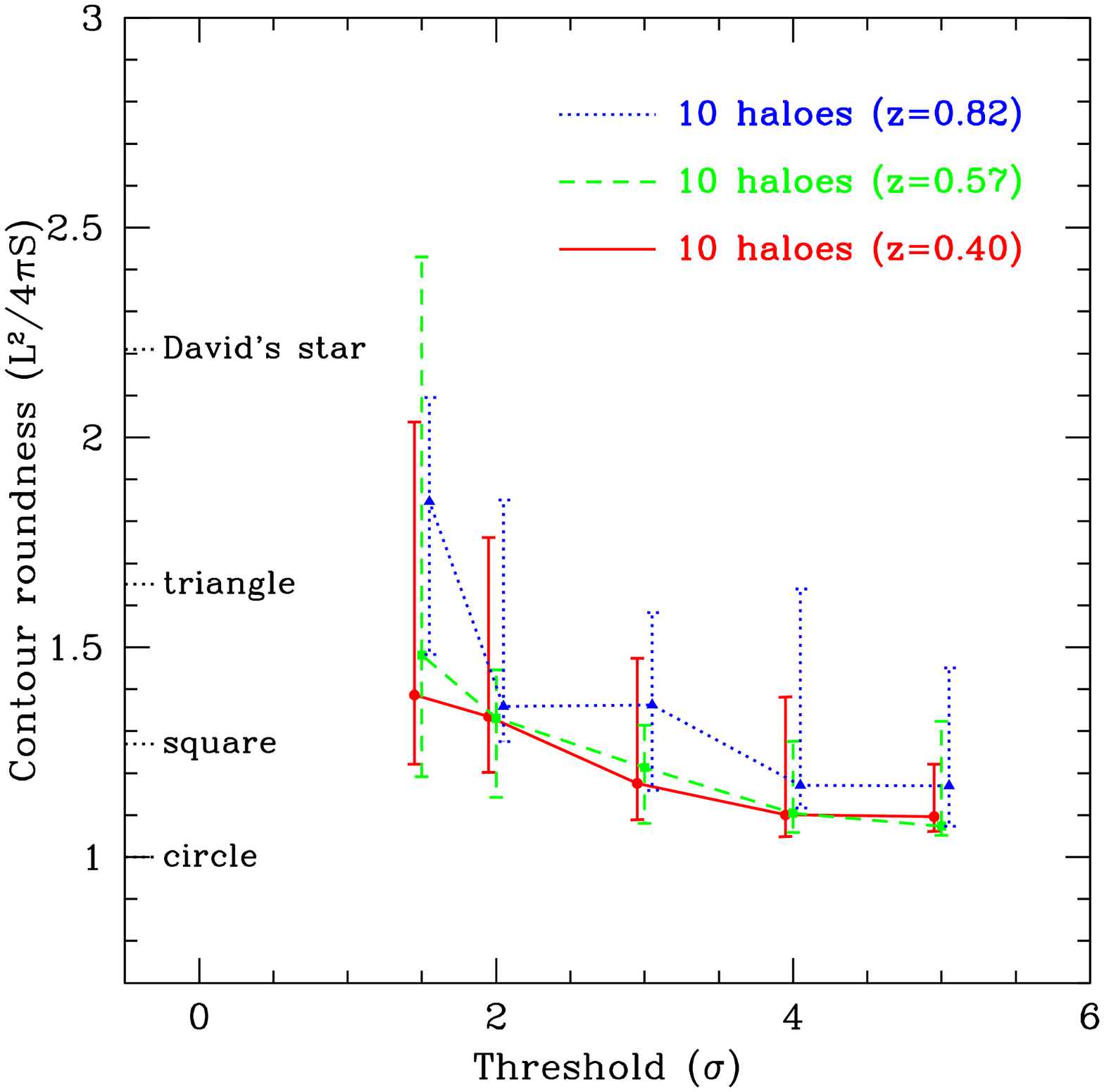}
\end{center}
\caption{
Roundness of iso-density contours ($L^2/4\pi S$) as a function of
the threshold of the contours, where $L$ is the
length of the contour and $S$ is the area surrounded by the contour.
Only the major clump near or at the cluster center (0,0) is used.
Observed data are plotted in the left panel, while the model predictions
are shown in the right panel.
The data points and the associated error-bars in the right panel
show the median and 16\%--84\% range of the distribution of 10 model halos
at each threshold, respectively.}
\label{fig:contour_shape}
\end{figure*}

\begin{figure}
\begin{center}
  \FigureFile(80mm,80mm){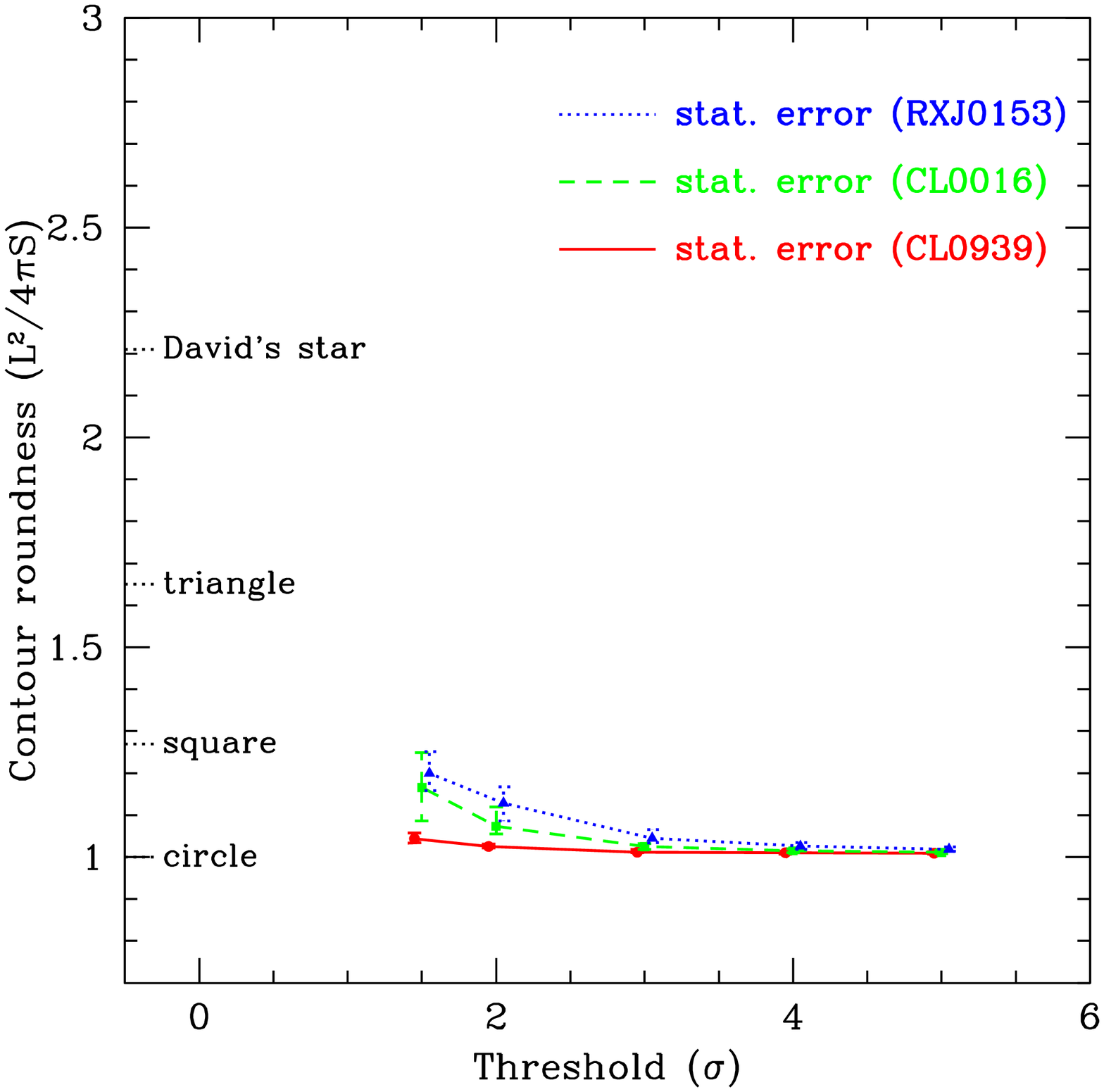}
\end{center}
\caption{
Statistical errors on the roundness index of iso-density contours
($L^2/4\pi S$) for the three clusters.
These were estimated from a Monte-Carlo simulation for each cluster
by generating artificial clusters that have perfectly round galaxy
distribution with $\Sigma\propto r^{-2}$
within 2~Mpc radius (flat distribution within 0.2~Mpc). On top of them
we add field contamination randomly over the entire field.
The number of the generated cluster galaxies within the 2~Mpc radius
and that of the generated field galaxies are equal to the statistically
estimated numbers from the observation, assuming that a low-density field
in the image corresponds to a general field.
We then measured the shape indices skewed by the contamination.
The data points and the associated error-bars 
show the median and 16\%--84\% range of the distribution of 10 Monte-Carlo runs.
}
\label{fig:contour_error}
\end{figure}

The roundness index of an iso-density contour is defined by
\begin{equation}
C=L^2/4\pi S
\end{equation}
in analogy to Schmalzing et al.\ (1999),
where $L$ is the length of the contour, and $S$ is the area surrounded
by the contour.  This index represents the roundness of the contour
and a perfect circle gives unity.  A square and an equilateral
triangle give shape indices of 1.27 and 1.65, respectively.
David's star, where an equilateral triangle is placed up-side down
on top of another triangle, gives $C=2.21$.
The larger is the index, the larger is the deviation from a circle.

The $C$ indices are measured for each iso-density contour of the phot-$z$
members drawn on the major clump at or near the cluster center
(figure~\ref{fig:map_lss}).  We plot them as a
function of the threshold of the contour ($\sigma$) in the left panel of
figure~\ref{fig:contour_shape} for the three clusters at different redshifts.
The general trend seen in all clusters is that the inner (higher) contours
are closer to circles, but the outer (lower) contours show larger deviations
from circles, indicating more complicated structures that reflect the filaments
coming out from the main bodies, as is evident from the 2-D distribution
in figures~\ref{fig:map_center} and \ref{fig:map_lss}.

The observed clusters after the photometric selection still contain
non-negligible field contamination at the lower contours, and it can skew
the intrinsic shape of cluster galaxy distribution.
To estimate this effect, we generated a perfectly round cluster
and added field contamination randomly over the entire field
by a Monte-Carlo simulation.  We then measured the shape indices, $C$, to
see how much the round galaxy distribution was skewd.
However, as shown in figure~\ref{fig:contour_error}, such a statistical
effect is found to be small.
On the other hand, if the intrinsic distribution of cluster galaxies are
far from round, the uniform (random) field contamination would make the
contours rounder and the measured $C$-indices would be lower limits.
In any case, we conclude that the statistical error in measuring
the contour shapes is not so large as to affect the trends we see in
figure~\ref{fig:contour_shape}.

CL~0939 and CL~0016 have similar shape indices, but the highest redshift cluster,
RX~J0153, has significantly higher values than the two lower redshift clusters
at all contour levels. This might indicate a possible evolutionary effect,
but we cannot make any general remarks with only three clusters and one each
at each epoch. We defer any conclusion until we complete observations
of most of the 15 clusters and make a statistical analysis on this plane.
We compare the shape indices between the observed clusters and
the simulated clusters in the right panel of figure~\ref{fig:contour_shape}
(see subsection 3.5).

\begin{figure*}
\begin{center}
  \FigureFile(80mm,80mm){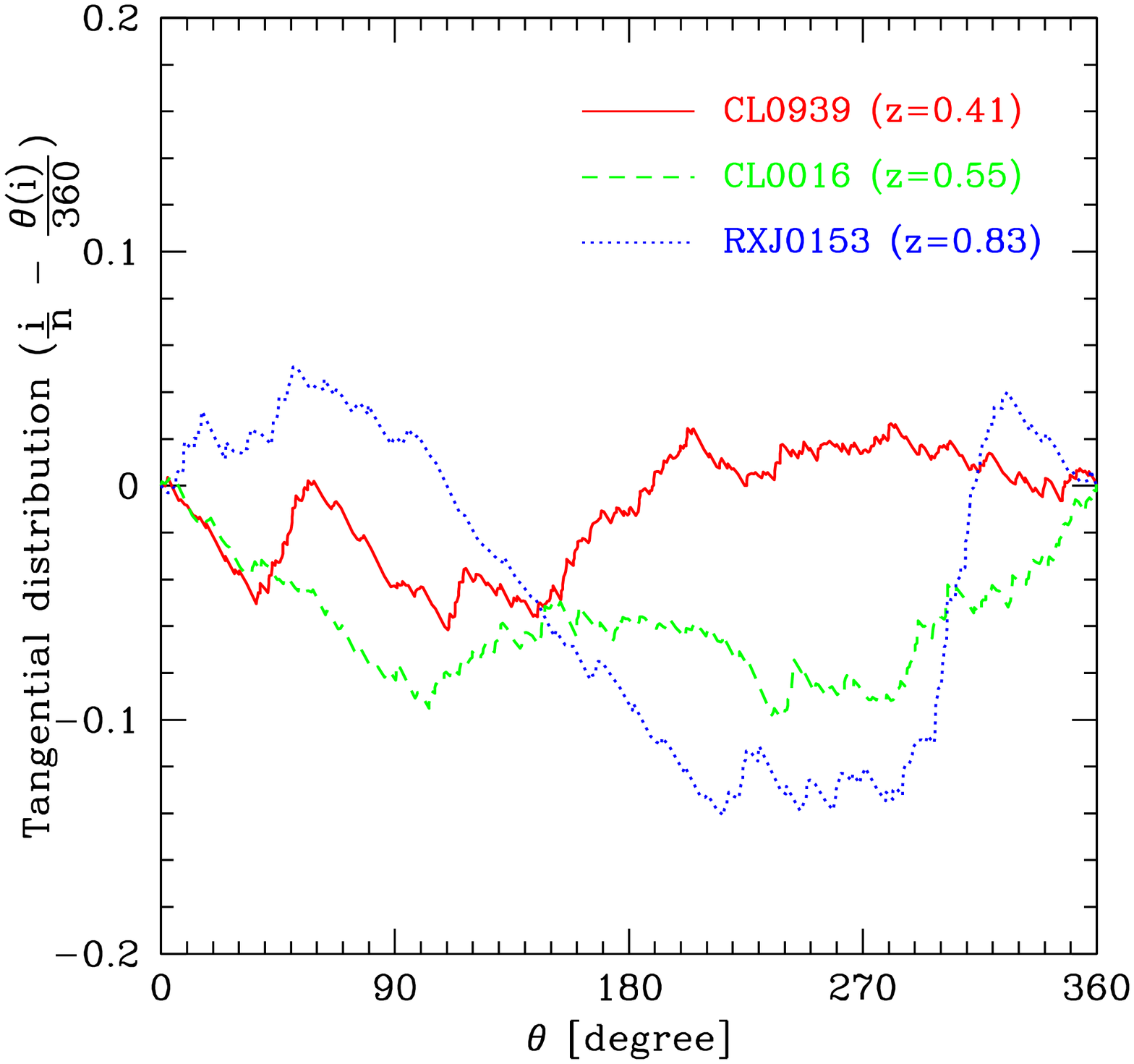}
  \hskip 0.5cm
  \FigureFile(80mm,80mm){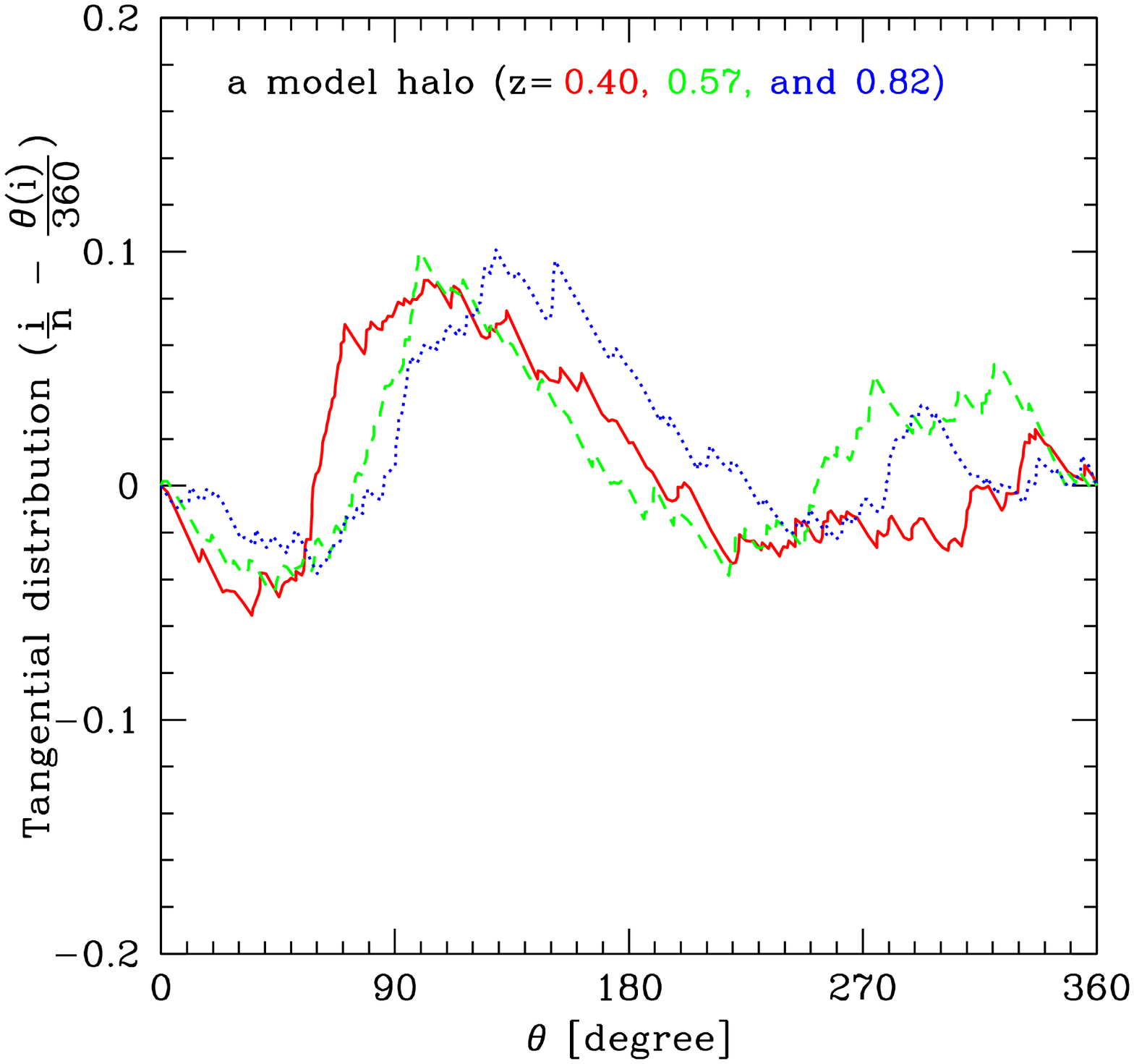}
\end{center}
\caption{
Tangential galaxy distribution within the radial range
of $0.3<R_{\rm c}<1.5$~Mpc centerd on the positions of
($\Delta$ R.A., $\Delta$ Dec.) =
(0.\hspace{-2pt}$'$4, 0.\hspace{-2pt}$'$2), (0$'$, 0$'$),
 and (0.\hspace{-2pt}$'$4, 0.\hspace{-2pt}$'$6),
for CL~0939, CL~0016, and RX~J0153 clusters, respectively.
The galaxies within the above radial range (total number is $n$) are sorted
in order of $\theta$, and the amplitude at the $i$-th galaxy at $\theta(i)$
is given by $\frac{\displaystyle i}{\displaystyle n}-\frac{\displaystyle
\theta(i)}{\displaystyle 360^{\circ}}$. 
A straight line at zero would mean the uniform
distribution of galaxies in the tangential direction
(equal spacing in $\theta$).
The decrement and increment of the curves reflect the
lower and higher densities compared to the averaged density, respectively
(left panel).
An example of our model cluster (most massive halo in our simulation)
is shown in the right panel.
}
\label{fig:filament}
\end{figure*}

\begin{figure*}
\begin{center}
  \FigureFile(80mm,80mm){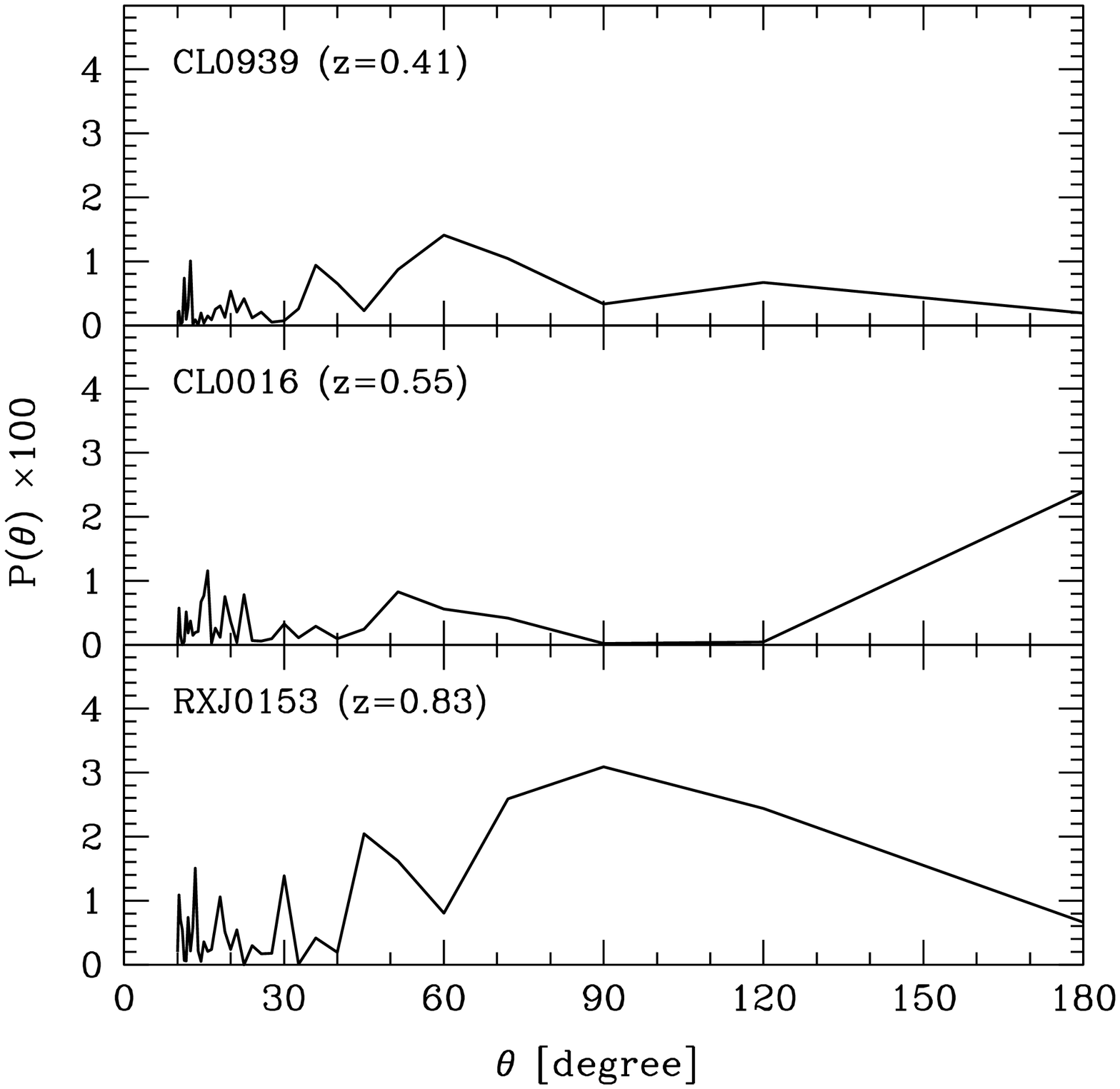}
  \hskip 0.5cm
  \FigureFile(80mm,80mm){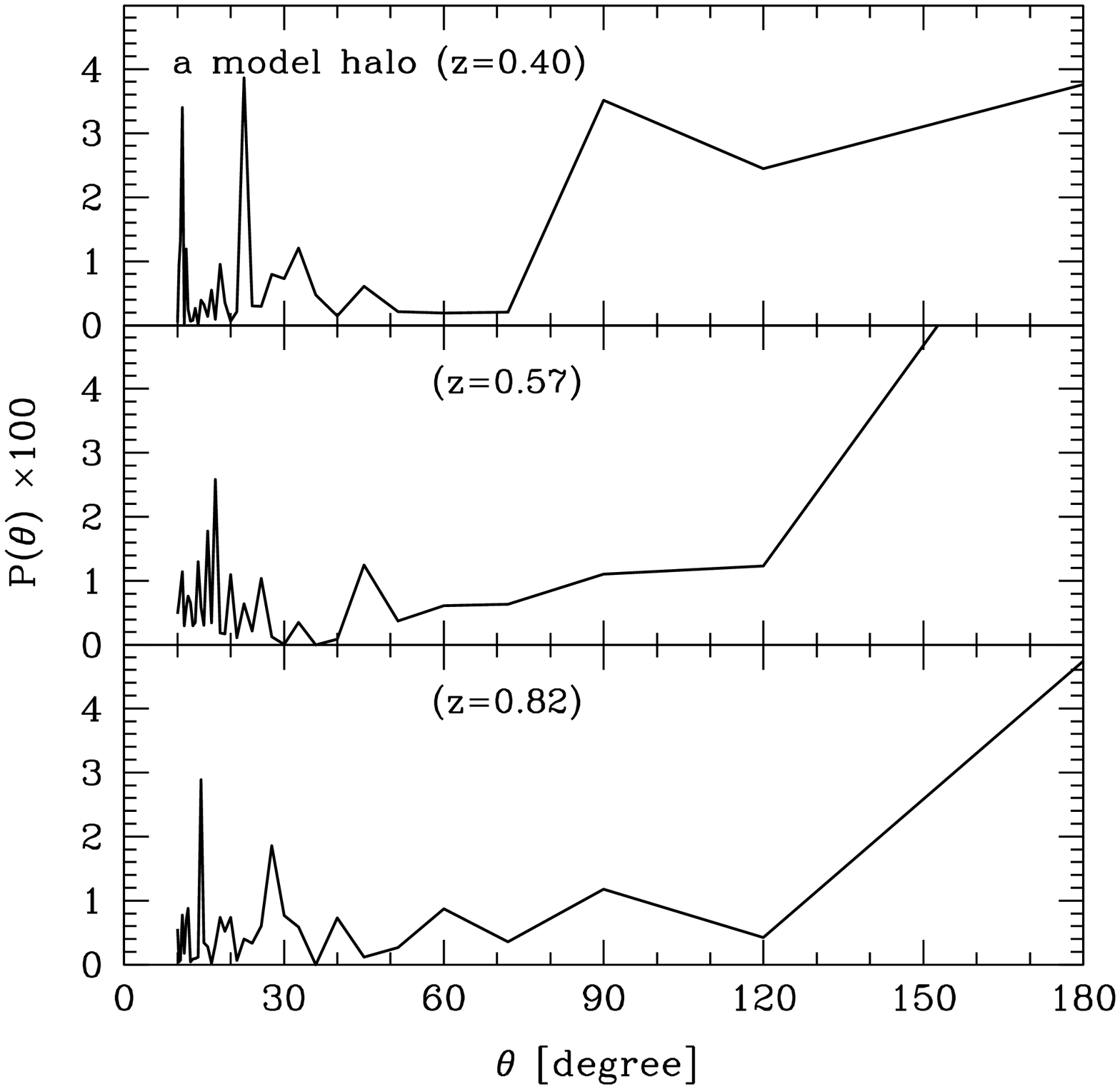}
\end{center}
\caption{
Normalized power of the Fourier expansion of the galaxy
distribution around the averaged density in the tangential direction,
within the radial range of $0.3<R_{\rm c}<1.5$~Mpc centerd on the positions of
($\Delta$ R.A., $\Delta$ Dec.) =
(0.\hspace{-2pt}$'$4, 0.\hspace{-2pt}$'$2), (0$'$, 0$'$),
and (0.\hspace{-2pt}$'$4, 0.\hspace{-2pt}$'$6), for CL~0939, CL~0016, and RX~J0153 clusters,
respectively (left panel).
An example of our model cluster (most massive halo in our simulation)
is shown in the right panel.
}
\label{fig:fourier}
\end{figure*}

In order to quantify the visual impression of filametary structures
of the observed distant clusters, we also plotted the tangential distribution
of galaxies (figure~\ref{fig:filament}) and calculated the power of the Fourier
expansion of the galaxy distribution around the averaged density in the
tangential direction (figure~\ref{fig:fourier}),
within the radial range of $0.3<R_{\rm c}<1.5$~Mpc of each cluster.
The galaxies within this radial range (total number is $n$) are sorted
in order of $\theta$, and the amplitude ($A$) in figure~\ref{fig:filament}
at the $i$-th galaxy at $\theta_i$ is defined as
\begin{equation}
A(\theta_i)=\frac{\displaystyle i}{\displaystyle n}-\frac{\displaystyle\theta_i}{\displaystyle 360^{\circ}}.
\end{equation}
Also, the normalized Fourier power, $P(\theta)$, in figure~\ref{fig:fourier}
is defined as
\begin{equation}
P(\theta)=\left[\frac{1}{n}\sum_{i=1}^{n} \cos k \theta_i\right]^2 + \left[\frac{1}{n}\sum_{i=1}^{n} \sin k \theta_i\right]^2,
\end{equation}
where $\theta=\frac{\displaystyle 360^{\circ}}{\displaystyle k}$ ($k=$1,2,3,...,36).
If the galaxies are distributed with equal spacing in the $\theta$ direction,
$A=P(\theta)=0$ (constant), by definition.
The results for the observed clusters are shown in the left panels,
while an example of our
model cluster (most massive halo in our simulation) is shown in the right
panels for comparison (see subsection 3.5).

The tangential distribution of galaxies
in the CL~0939, CL~0016 and RX~J0153 clusters show high powers at 60$^{\circ}$,
180$^{\circ}$, and 90$^{\circ}$, respectively, which reflect the visual
impression of the structures of these clusters: namely, many ($\sim6$)
filaments coming out from the core of CL~0939 (figure~\ref{fig:map_lss}),
the global linear structure in the NE--SW direction of CL~0016
(figure~\ref{fig:map_center}), and the `$T$'-shape structure in the core of 
RX~J0153 (figure~\ref{fig:map_center}).

\subsection{Comparison with Numerical Simulations}
\label{sec:simulation}

We compare our results with $\nu$GC simulation of galaxy formation
(Nagashima et al.\ 2005)
which is based on the CDM based $N$-body simulations
(Yahagi et al.\ 2004)
and on the Mitaka semi-analytic model of galaxy formation
(Nagashima, Yoshii 2004).
The simulation volume was a 100$\times$100$\times$100~Mpc$^3$ cube (comoving),
and the number of particles used was 512$^3$.  The mass of each particle
was 3.04$\times$10$^8$M$_{\odot}$.
The adopted cosmological parameters were ($h_{70}$, $\Omega_m$,
$\Omega_{\Lambda}$)=(1.0, 0.3, 0.7), $\Omega_{\rm b}$=0.04,
and $\sigma_8$=1.
The power spectrum given by Bardeen et al.\ (1986) was used.
The model successfully reproduced many observational quantities,
such as the luminosity functions, gas fractions, and faint galaxy counts
(Nagashima et al.\ 2005).

Halos were identified from the simulation data by applying the
friends-of-friends algorithm with a linking length of 20~\%
of a mean separation of particles at each redshift.
We selected those halos that contain equal to, or
more than 10 particles; hence, the mass resolution was
3$\times$10$^9$~$M_{\odot}$.
The masses of the halos were determined at each epoch within
the iso-density surface at which the mass density was 200
times higher than the mean density of the Universe.
We used 10 of the most massive halos at $z=0.41$, 0.57, and 0.82,
respectively, for comparing with CL~0939, CL~0016 and RX~J0153.
The masses of the 10 selected cluster halos ranges from
5.5$\times$10$^{14}$$M_{\odot}$ (heaviest) to
9.5$\times$10$^{13}$$M_{\odot}$ (lightest) at $z=0.82$,
and these masses evolved by a factor of 1.3--2 by the present-day,
except for the halos incorporated into more massive halos.
The masses of the observed three clusters inside the 1~Mpc radii
are in the range of 2--5$\times$10$^{14}$$M_{\odot}$
(De Filippis et al.\ 2003; Worrall, Birkinshaw 2003; Huo et al.\ 2003),
spanning the range of the model cluster masses.
We note that the mass range of the model halos extends to slightly
smaller mass compared to that of the observed clusters due to the 
limitation of the simulation box.  Therefore, we may need a larger
simulation to make a truly fair comparison.
The right panels in figures~\ref{fig:map_lss}, \ref{fig:filament},
and \ref{fig:fourier} show the 2-D projection of the
most massive halo in the simulation box, as well as its tangential distribution and
its Fourier expansion, respectively.  The filamentary and clumpy
structures of the model clusters look very similar to those of the observed
clusters.

In order to make a quantitative comparison,
the contour shapes, $C$, were calculated for these 10 model clusters at each
epoch ($z$=0.40, 0.57, and 0.82), exactly in the same manner as for the observed
clusters, where we used a smoothing length of 0.2~Mpc and bin size of 0.1~Mpc
in the physical (proper) scale.
The model clusters were projected to 2-D in this calculation over a depth of
30~Mpc (comoving).
The shape indices of the model clusters are shown in the right panel of
figure~\ref{fig:contour_shape}.  The error-bars show the 16\%--84\% range
of the distribution of 10 halos at each threshold.
The model values are in good agreement with the observed ones for
$z\sim 0.4$ and $z\sim 0.55$, and
the general trend that the contour shapes deviate from circles with decreasing
density is seen in the model clusters as well.
The higher redshift clusters at $z=0.82$ in the model, however, show much
lower shape indices than observed, although they are still systematically
slightly larger than those of the lower redshift clusters in the model.
The evolution of the contour shapes seems to be weaker in the model than
observed, although the number of observed data points should be increased
to properly discuss the evolution.

\section{Summary}
\label{sec:summary}

We have started a distant cluster project, PISCES, on Subaru.
Its general concepts and first results concerning the cluster structures
of three clusters are presented in this paper.
We have found spectacular large-scale
structures across $\sim$30$'$ scales around the well-studied cluster cores
of the three clusters, i.e., CL~0939 ($z=0.41$), CL~0016 ($z=0.55$),
and RX~J0153 ($z=0.83$) based on multi-color wide-field
imaging with Suprime-Cam on Subaru.
These structures are characterized by filaments coming out from the core and
bridging some clumps along the filaments.
These overall large-scale structures are in qualitative agreement with
the numerical simulations, which predict an infall of clumps into 
the cluster cores travelling along the filaments during the course
of cluster-scale assembly.

Moreover, we have taken a closer look at near or inside the cluster cores,
and have identified complicated substructures that seem to be connected
to the outer structures, and hence suggesting a recent accretion of
matter from the surrounding filaments.

The substructures of the main clusters were compared with the numerical
simulation ($N$-body + semi-analytic model) applying the deviation of
shapes of the iso-density contours from perfect circles.
A quantitative agreement between the observation and the model
was found at $z\sim0.4$ and $z\sim0.55$, although the model predicts less
evolution at $z\sim0.83$.  We do not know, however, whether this is a true
evolutionary effect, or just an exception.
To increase the sample of clusters is obviously needed to quantify
any evolutionary signatures of clusters by distinguishing from
the individualities of clusters.
Also, we need to extend the simulation box while keeping its resolution 
to have a statistically fair sample of rich clusters, so that we can directly
compare them with our observed clusters.

The optical substructures in the inner regions of clusters and their
neighbouring regions were also compared to the X-ray maps reproduced from
the archived XMM data,
and were found to have a good correlation with the X-ray extended
emissions.  This indicates that most of these sub-systems near the cluster
cores are physically bound objects, rather than chance projections
along the lines of sight.
We are therefore likely to be witnessing a hierarchical growth of
clusters.

%
%
%
We acknowledge Drs.\ Richard Bower, Laurence Jones and John Blakeslee
for helpful discussions.
We also thank Drs.\ K.\ Saigo and K.\ Ichiki at NAOJ for helping
us to develop the software used in the analysis.
This work is financially supported in part by a Grant-in-Aid for the
Scientific Research (No.\ 15740126, 14102004, 16340053, and 16540223)
by the Ministry of Education, Culture, Sports, Science and Technology.
M.T., N.M., and T.O. acknowledge the JSPS research fellowship.

\end{document}